\documentclass[aps,twocolumn,showpacs,10pt]{revtex4-1}
\usepackage[utf8]{inputenc}
\usepackage{graphicx,color,amsmath}
\usepackage{amssymb}
\usepackage{amsthm}
\usepackage[bookmarks]{hyperref}
\usepackage{umoline}

\newcommand{\be}{\begin{equation}}
\newcommand{\ee}{\end{equation}}
\definecolor{mygreen}{rgb}{0,0.5,0}
\definecolor{myblue}{rgb}{0,0,0.75}
\definecolor{mymagenta}{cmyk}{0,1,0,0.12}

\begin{document}
\title{Dynamical topological order parameters far from equilibrium}
\author{Jan Carl Budich}
\author{Markus Heyl}
\affiliation{Institute for Quantum Optics and Quantum Information of the Austrian Academy of Sciences, 6020 Innsbruck, Austria}
\affiliation{Institute for Theoretical Physics, University of Innsbruck, 6020 Innsbruck, Austria}
\date{\today}
\begin{abstract}
We introduce a topological quantum number -- coined dynamical topological order parameter (DTOP) -- that is dynamically defined in the real-time evolution of a quantum many-body system and represented by a momentum space winding number of the Pancharatnam geometric phase. Our construction goes conceptually beyond the standard notion of topological invariants characterizing the wave-function of a system, which are constants of motion under coherent time evolution. In particular, we show that the DTOP can change its integer value at discrete times where so called dynamical quantum phase transitions occur, thus serving as a dynamical analog of an order parameter. Interestingly, studying quantum quenches in one-dimensional two-banded Bogoliubov de\,Gennes models, we find that the DTOP is capable of resolving if the topology of the system Hamiltonian has changed over the quench. Furthermore, we investigate the relation of the DTOP to the dynamics of the string order parameter that characterizes the topology of such systems in thermal equilibrium.
\end{abstract}

\maketitle
\section{Introduction}
In the theory of phase transitions~\cite{Fisher1967da,Sachdev2011}, the phases separated by a transition can be qualitatively distinguished in terms of order parameters which can either be local, following the conventional Ginzburg-Landau paradigm, or can be of topological nature. Regarding the latter, a complete classification for the ground states of gapped fermionic band structures in terms of global topological invariants has been achieved \cite{Schnyder2008,Kitaev2009,Ryu2010}. Out of thermal equilibrium, however, the current understanding of such systems is much less systematic. In particular, standard topological invariants \cite{Schnyder2008,Kitaev2009,Ryu2010} associated with the instantaneous wave function are constants of motion within unitary time evolution and can therefore not capture genuine non-equilibrium effects.
In a non-equilibrium context, several recent studies \cite{FosterPRB,FosterPRL,RigolPreprint,CaioPreprint} have discussed modifications to the bulk-boundary correspondence in non-equilibrium settings, indicating that the dynamical formation of edge states follows the instantaneous Hamiltonian and the single particle Green's function, respectively.

In this work, we construct a bulk topological quantum number -- coined dynamical topological order parameter (DTOP) -- which is dynamically defined: Going conceptually beyond the classification of topological phases in thermal equilibrium, it characterizes topological properties of the real-time dynamics rather than of the instantaneous wave function or the instantaneous Hamiltonian. In particular, we show in what sense the DTOP, somewhat analogous to order parameters at conventional phase transitions, distinguishes two phases separated by a nonequilibrium transition occurring in the coherent time evolution of a quantum system. Specifically, we find that the DTOP changes its value at a dynamical quantum phase transition (DQPT) \cite{Heyl2013a} which appears as a non-analytic behavior at critical times of the Loschmidt amplitude
\be
\mathcal G(t)=\langle \psi\vert \psi(t)\rangle=\langle \psi\rvert \text{e}^{-iHt}\lvert \psi\rangle,
\label{eqn:loschmidtAmplitude}
\ee
where $|\psi\rangle$ denotes the initial state and $H$ the Hamiltonian governing the nonequilibrium quantum real-time evolution. 
Intriguing properties of DQPTs have been identified in a variety of different systems~\cite{Heyl2013a,Pollman,Obuchi,Heyl2014,Fagotti2013,Karrasch2013,Kriel2014,Andraschko2014,Hickey2014,Canovi2014fo,Vajna2014,VanjaDoraPreprint,Konik2015,Heyl2015}. 
While Loschmidt amplitudes bear a formal similarity to equilibrium partition functions~\cite{Heyl2013a}, a dynamical analogue of an order parameter, that physically distinguishes the time intervals separated by a DQPT, has not yet been identified. Here, we show how to construct such a dynamical analog for DQPTs by studying quantum quenches in two-banded  Bogoliubov de\,Gennes models such as the Kitaev chain \cite{Kitaev2001}. Notice that notions of dynamical transitions occuring out of equilibrum have also been introduced in different contexts \cite{DynamicalTransitions}. In the following, however, we will be referring to the definition made in Ref. \cite{Heyl2013a} in terms of Loschmidt amplitudes (see Eq.~(\ref{eqn:loschmidtAmplitude})).

\begin{figure*}[htp]
\centering
\includegraphics[width=0.9\textwidth]{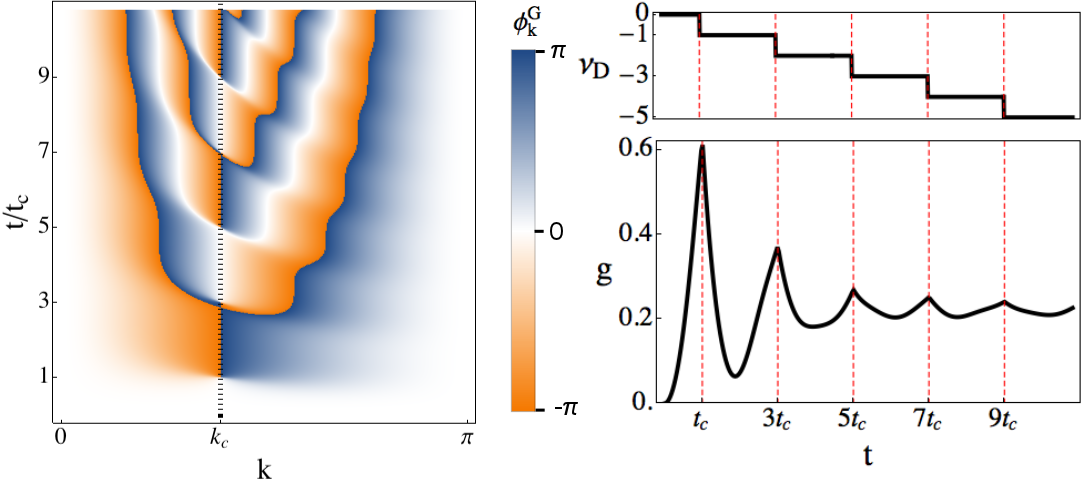}
\caption{\label{fig:one} (color online) Left panel: Color plot of the Pancharatnam geometric phase $\phi^G_k(t)$, see Eq.~(\ref{eqn:PGP}), for chemical-potential quenches $\mu = 0 \to \mu = 3$ in the Kitaev chain as a function of lattice momentum and time. Time is measured in units of the critical time $t_c$ where the first dynamical quantum phase transition (DQPT) occurs. The critical momentum $k_c$ at which non-analyticities occur is marked with a black dotted line. Right panel: Rate function $g(t)=-N^{-1}\mathrm{Re}\log[|\mathcal G(t)|^2]$ of the Loschmidt echo $\mathcal{L}(t)=|\mathcal G(t)|^2$, see Eq.~(\ref{eqn:loschmidtAmplitude}), and the dynamical topological order parameter $\nu_D(t)$, see Eq.~(\ref{eqn:dtop}), as a function of time. The real-time non-analyticities in $g(t)$, occuring at odd multiples of $t_c$ (red dashed lines), define the DQPTs. $\nu_D(t)$ changes its value only at the DQPTs thus serving as a dynamical order parameter.}
\end{figure*}

DQPTs occur whenever the time-evolved state $\lvert \psi(t)\rangle$ becomes orthogonal to the initial state vector $\lvert \psi\rangle$. This can be formally understood
from the concept of partition function zeros~\cite{Yang1952ez,Lee1952vl,Fisher1967da}, which occur as so called {\emph{Fisher zeros}} in the context of Loschmidt amplitudes~\cite{Heyl2013a}. Concerning the quest to identify dynamical order parameters for DQPTs, this direct relation between DQPTs and wave function orthogonalities guides our intuition towards looking for an observable quantity that is smoothly defined for non-orthogonal state vectors and thus only allowed to behave discontinuously at critical times. The Pancharatnam geometrical phase (PGP) \cite{Pphase,PphasePRL} is precisely such a quantity. It was originally introduced \cite{Pphase} to define a relative phase for light beams with non-orthogonal polarization and has later been generalized to extend the notion of Berry's geometric phase \cite{Berry,Simon} to general time evolution with non-orthogonal initial and final states, in particular allowing for non-adiabatic \cite{AAphase} and non-cyclic \cite{PphasePRL} dynamics. The geometric background of this construction is that a non-cyclic evolution can be augmented to a cyclic path in a unique way only if the two end states are {\emph{non-orthogonal}}, namely by going back from the final to the initial state along a geodesic in projective Hilbert space.

The DTOP introduced in this work is a momentum-space winding number of the PGP which serves as a dynamical analog of a \emph{topological} order parameter in two-banded  Bogoliubov de\,Gennes models undergoing a DQPT after a quantum quench, i.e., a sudden change in the band structure parameters. 
We show that the integer-valued DTOP can change its value only at DQPTs and, moreover, that this allows us to dynamically resolve how the topology of the underlying Hamiltonian has changed during the quench. Our construction relies on the presence of a so-called particle hole symmetry (PHS), i.e., a spectral constraint imposed by an antiunitary operator $\mathcal C,~\mathcal C^2=1$ which anti-commutes with the system Hamiltonian. In  Bogoliubov de\,Gennes models, such a constraint is naturally imposed by the fermionic algebra to the Nambu spinor representation of the Hamiltonian \cite{AltlandZirnbauer}. We illustrate our construction by studying quenches in several models including, e.g. the Kitaev chain \cite{Kitaev2001}.\\

{\emph{Outline. }}The remainder of this article is organized as follows. Section \ref{sec:model} is devoted to the definition of our model system as well as the general analysis of its quench dynamics, in particular the occurrence of DQPTs. In Section \ref{sec:DefDTOP}, we construct the DTOP which is at the heart of this study and discuss its basic properties. Thereafter, in Section \ref{sec:simulations}, we present benchmark simulations, showing how the DTOP characterizes the dynamics of several 1D Bogoliubov de\,Gennes models. The relation to the familiar string order parameter which hallmarks different phases in equilibrium is investigated in Section \ref{sec:stringorder}. Finally, in Section \ref{sec:discussion}, we outline how the physics discussed in this manuscript, in particular the DTOP, could be experimentally observed, and put our present work into a broader context by discussing its relation complementary approaches.

\section{Model and Dynamics}
\label{sec:model}
\subsection{Underlying Hamiltonian}
The dynamical properties we are concerned with here are generated by gapped free fermionic two-banded  Bogoliubov de\,Gennes models in 1D without requiring further symmetries, i.e., in symmetry class D \cite{AltlandZirnbauer}. We denote the Nambu pseudo-spin by $\tau$ and choose the convention $\mathcal C=\tau_1 K$ for the PHS operation, where $K$ denotes the complex conjugation. Assuming a unit lattice constant, the first Brillouin zone is the circle resulting from the interval $[-\pi,\pi]$ by identification of its end points. The Bloch Hamiltonian is then of the form
\begin{align}
H(k)=\vec d(k)\cdot \vec \tau=\sum_{j=1}^3 d^j(k)\tau_j
\label{eqn:hamform}
\end{align}
and satisfies the spectral PHS constraint
\begin{align}
\tau_1 H(k) \tau_1=-H^*(-k).
\label{eqn:phs}
\end{align}
As a consequence of Eq. (\ref{eqn:phs}), $d^1(k)$ and $d^2(k)$ must be odd functions of the lattice momentum $k$, while $d^3(k)$ must be even.
Eq. (\ref{eqn:phs}) is local in momentum at the two real lattice momenta $k_{\mathbb R}=0,\pi$ which satisfy $k=-k ~(\text{mod} 2\pi)$ and where both $d^1$ and $d^2$ need to vanish such that
\begin{align}
H(k_{\mathbb R})=d^3(k_{\mathbb R}) \tau_3,~k_{\mathbb R}\in \left\{0,\pi\right\}.
\label{eqn:hreal}
\end{align}
With the unit vector $\hat d(k)=\vec d(k)/\lvert \vec d(k)\rvert$, the geometrical interpretation of Eq. (\ref{eqn:hreal}) is that $\hat d$ is pinned to the poles of the Bloch sphere at the real momenta which plays a crucial role in the following. There are two topologically inequivalent classes of such Hamiltonians \cite{Kitaev2001}, here simply distinguished by the sign of $d^3(0)d^3(\pi)$ which becomes negative for the non-trivial topological phase.

\subsection{Quench dynamics and DQPTs} 
We study nonequilibrium quantum real-time evolution and DQPTs induced by a quantum quench. The system is prepared in the ground state $\lvert\psi\rangle$ of an initial Hamiltonian $H_i(k) = \vec d_i(k) \cdot \vec \tau$. At time $t=0$, a parameter will be switched suddenly within the set of models in Eq.~(\ref{eqn:hamform}) resulting in a sudden change $\vec d_i(k) \mapsto \vec d_f(k)$.
We assume that the system initially occupies the lower Bloch band of $H_i$. The associated lower Bloch states are denoted by $\lvert u^{i-}_k\rangle$ such that the initial state $\lvert \psi \rangle$ is a Slater determinant of all lower band Bloch states. Since lattice translation invariance is maintained at all times, the dynamics of the system can be considered separately for every lattice momentum $k$. Explicitly, we get
\begin{align}
\lvert \psi_k(t)\rangle=\text{e}^{i \epsilon_k^f t} g_k \lvert u^{f-}_k\rangle + \text{e}^{-i \epsilon_k^f t} e_k \lvert u^{f+}_k\rangle,
\label{eqn:gendyn}
\end{align}
where $\pm\epsilon_k^f=\pm \lvert \vec d_f(k)\rvert$ denotes the energy eigenvalues of $H_f(k)$, $\lvert u^{f\pm}_k\rangle$ its Bloch states, and $g_k=\langle u^{f-}_k\vert u^{i-}_k\rangle, e_k=\langle u^{f+}_k\vert u^{i-}_k\rangle$ with $\lvert g_k\rvert^2=\frac{1}{2}(1+\hat d_i(k)\cdot \hat d_f(k)),~\lvert e_k\rvert^2=\frac{1}{2}(1-\hat d_i(k)\cdot \hat d_f(k))$ are expansion coefficients of the initial lower Bloch state in the new Bloch states after the quench. For the geometric interpretation of our later results, it is helpful to consider the vector $\hat d_i(k)$ as a reference direction, say pointing to the south pole of a Bloch sphere defined at every momentum, and to consider the direction of $\hat d_f(k)$ relative to this reference. We refer to this construction as the {\emph{relative Bloch sphere}} in the following.

Recently, DQPTs in topological systems satisfying Eq.~(\ref{eqn:hamform}) as well as in related spin chains have been identified~\cite{Heyl2013a,Vajna2014,VanjaDoraPreprint}. DQPTs are caused by Fisher zeros~\cite{Heyl2013a} where for a momentum $k_c$ the overlap $\langle u_{k_c}^{i-} | \psi_{k_c}(t_c)\rangle=0$ vanishes at a time $t_c$. Here, this can only happen if
\begin{align}
\lvert g_{k_c}\rvert^2= \lvert e_{k_c}\rvert^2~\text{and}~t_{c,n}= \frac{(2 n -1) \pi}{2 \epsilon_{k_c}^f},~n\in \mathbb N.
\label{eqn:zerocond}
\end{align}
Fisher zeros and DQPTs hence occur at momenta where the initial lower Bloch state is an equal weight superposition of the final Bloch states, i.e. at $\hat d_i(k_c)\cdot \hat d_f(k_c)=0$ marking the equator of the relative Bloch sphere, whereas the critical time is determined by the spectrum $\epsilon_k^f$ of the final Hamiltonian.\\

\section{Construction of a dynamical topological quantum number}
\label{sec:DefDTOP}
\subsection{Pancharatnam geometric phase} 

In order to define the PGP at lattice momentum $k$, let us decompose the Loschmidt amplitude $\mathcal{G}(t) = \prod_{k>0} \mathcal{G}_k(t)$ with
\begin{align}
\mathcal{G}_k(t)=\langle u_{k}^{i-} | \psi_{k}(t)\rangle =r_k(t)\text{e}^{i\phi_k(t)}.
\label{eqn:lopolar}
\end{align}
and $r_k(t)$, $\phi_k(t)$ its polar coordinates. The phase $\phi_k(t)$ contains a purely geometric and gauge-invariant component
\be
	\phi_k^\text{G}(t) = \phi_k(t) - \phi_k^\text{dyn}(t)\,,
	\label{eqn:PGP}
\ee
obtained by subtracting the dynamical phase $
\phi_k^{\text{dyn}}(t)=-\int_0^t\, \text{d}s \langle \psi_k(s)\rvert H_f \lvert \psi_k(s)\rangle=\epsilon_k t (\lvert g_k\rvert^2-\lvert e_k\rvert^2)$. $\phi_k^\text{G}$ is the aforementioned PGP \cite{PphasePRL} that will be the central building block for the DTOP introduced in this work.
We stress that this definition of the PGP becomes singular at Fisher zeros as the total phase $\phi_k(t)$ in Eq. (\ref{eqn:lopolar}) is ill-defined at critical times.

\subsection{Definition of the DTOP}
Eq. (\ref{eqn:hreal}) implies that either $\lvert e_k\rvert^2=0$ and $\lvert g_k\rvert^2=1$ or vice versa at the real momenta $k_{\mathbb R}=0,\pi$. From Eq. (\ref{eqn:gendyn}), we directly conclude that $\phi_{k_{\mathbb R}}(t)=\phi_{k_{\mathbb R}}^{\text{dyn}}(t)$, i.e., that the PGP is pinned to zero at these special momenta. Thus, as far as the PGP is concerned, the interval $[0,\pi]$ between the real momenta can be endowed with the topology of the unit circle $S^1$ by identifying its end points. We refer to this periodic structure as the effective Brillouin zone (EBZ). We are now ready to define a DTOP in terms of the PGP as
\begin{align}
\nu_D(t)=\frac{1}{2\pi}\oint_0^\pi \frac{\partial \phi_k^{\text{G}}(t)}{\partial k}.
\label{eqn:dtop}
\end{align} 
$\nu_D(t)$ is the integer-quantized winding number of the PGP over the EBZ and is smoothly defined as a function of time in the absence of Fisher zeros. More formally,
$\nu_D(t)$ is a topological invariant distinguishing homotopically inequivalent mappings $\text{EBZ}\rightarrow U(1),~ k\mapsto \text{e}^{i\phi_k^{\text{G}}(t)}$ from the unit circle $S^1$ to itself. The definition of $\nu_D(t)$ in Eq. (\ref{eqn:dtop}) and its subsequent further interpretation are the main results of our present work.

Since DQPTs can only occur at points in time where Fisher zeros are present, $\nu_D(t)$ must be constant in time intervals between DQPTs as it cannot smoothly change its integer value. But does $\nu_D$ change its value at every DQPT?
We answer this question in the affirmative implying that $\nu_D(t)$ can serve as an order parameter for the studied DQPTs. We find that, quite remarkably, the change in the DTOP $\Delta \nu_D(t_c)$ in the vicinity of a critical time $t_c$ can be directly related to the sign of the slope $s_{k_c}=\left.(\partial_k \lvert e_k\rvert^2)\right |_{k_c}$ at the critical momentum as
\begin{align}
\Delta \nu_D(t_c) =\lim_{\tau\to0} \left[\nu_D(t_c+\tau)-\nu_D(t_c-\tau)\right]=\text{sgn}(s_{k_c})
\label{eqn:index}
\end{align}
which loosely resembles an index theorem. This result affords an intuitive geometric interpretation:
As pointed out before, critical momenta are located on the equator of the relative Bloch sphere. $\Delta \nu_D(t_c)$ is then directly related to whether $\hat d_f(k)$ traverses the equator of the relative Bloch sphere from the northern to the southern hemisphere ($\text{sgn}\left(s_{k_c}\right)=-1$) or from the southern to the northern hemisphere ($\text{sgn}\left(s_{k_c}\right)=1$) at the critical momentum.

To establish Eq. (\ref{eqn:index}), we first identify a fundamental dynamical symmetry of $\mathcal{G}_k(t)$ (see Eq. (\ref{eqn:lopolar})) at critical momenta $k_c$: From Eqs. (\ref{eqn:gendyn}-\ref{eqn:zerocond}), we conclude that $\mathcal{G}_{k_c}(t)\in \mathbb R$. Furthermore, the dynamical phase is zero due to $\lvert g_{k_c}\rvert^2=\lvert e_{k_c}\rvert^2$ such that
$\text{e}^{i\phi^G_{k_c}(t)}=\text{sgn}\left(\cos(\epsilon_{k_c}^f t)\right)$, i.e., the PGP is pinned to the real values $0,\pi$ at the critical momenta at all times. When passing through critical times, marked by $\cos(\epsilon_{k_c}^f t_c)=0$ , the sign of $\cos(\epsilon_{k_c}^f t)$ changes and the PGP jumps by $\pi$. Expanding $\partial_k \phi^G_k(t)$ 
around $k_c$ and $t_c$ to leading order, it is straightforward to prove Eq. (\ref{eqn:index}). We note that the connection between Fisher zeros and $\pi$ jumps of the PGP is generally valid beyond the scope of the present DTOP: When a complex function (the Loschmidt amplitude) goes through zero as function of a real parameter (time), its phase jumps by $\pi$. Since the dynamical phase is always continuous in time, this jump can only occur in the PGP.

\begin{figure*}[htp]
\centering
\includegraphics[width=0.9\textwidth]{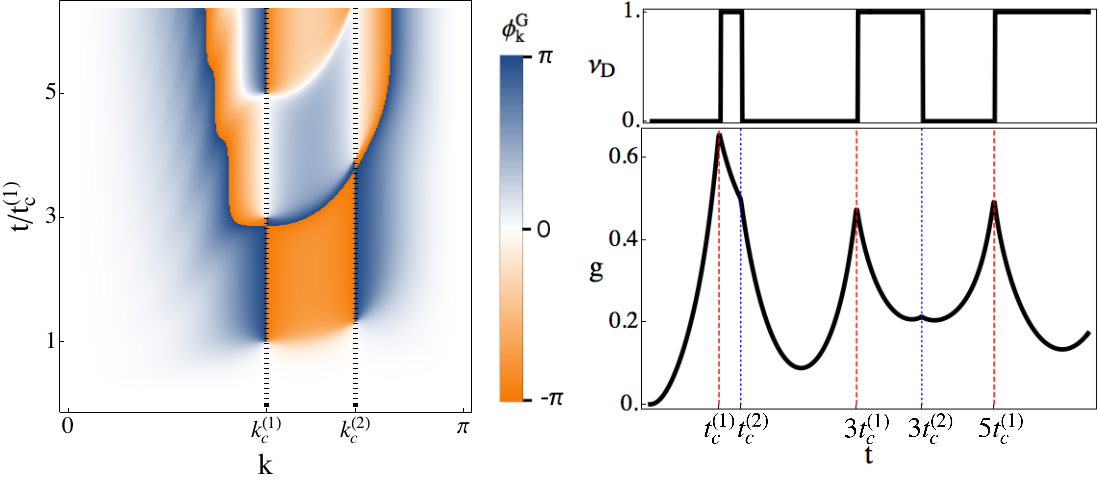}
\caption{\label{fig:two} Left panel: Color plot of the PGP $\phi^G_k(t)$ as a function of lattice momentum and time for $\lambda=1.3$. Time is measured in units of the first critical time $t_c^{(1)}=\pi/2$. The critical momenta $k_c^{(1)}$ and $k_c^{(2)}$ at which DQPTs appear as discontinuities in the PGP is marked with black dotted lines. Right panel: Rate function $g(t)$ and DTOP $\nu_D(t)$ as a function of time for $\lambda=1.3$. The DQPTs occur at odd multiples of the critical times $t_c^{(1)}$ and $t_c^{(2)}$.}
\end{figure*}

\section{Benchmarks simulations}
\label{sec:simulations}
We further investigate and illustrate the DTOP $\nu_D(t)$ with two benchmark examples. First, we study the Kitaev chain\cite{Kitaev2001}, a toy model for a proximity induced $p$-wave superconductor. The model Hamiltonian is of the form (\ref{eqn:hamform}) with $d^1(k)=0,~d^2(k)=\sin(k),~d^3(k)=\mu-\cos(k)$, where physically the $\cos(k)$ term represents the kinetic energy and the $\sin(k)$ term represents the $p$-wave pairing.
Here, we simulate a quench in the chemical potential from $\mu=0$ in the initial Hamiltonian $H_i$ to $\mu=3$ in the final Hamiltonian $H_f$. At the real momentum $k=0$ the $d^3$ component switches sign over the quench while at $k=\pi$ it does not, meaning that the topological phase of the Hamiltonian \cite{Kitaev2001} changes during the quench. Hence, $\lvert e_0\rvert^2=1$ while $\lvert e_\pi\rvert^2=0$.
Due to continuity, there must be a critical momentum $k_c$ in the interior of the EBZ where $\lvert e_{k_c}\rvert^2=\lvert g_{k_c}\rvert^2=\frac{1}{2}$ in agreement with Ref. \cite{VanjaDoraPreprint}. According to Eq. (\ref{eqn:zerocond}), this implies that DQPTs will occur at times $t_{c,n}=(2n-1)\pi/(2\epsilon_{k_c})=(2n-1) t_c,~n\in \mathbb N$. 
More specifically we find, $k_c=\arccos(1/3),~t_c = \pi/(4\sqrt{2})$ and $s_{k_c}=-1$, i.e., $\hat d_f(k)$ crosses the equator of the relative Bloch sphere in southern direction, corresponding to a change $\Delta(t_{c,n})=-1$ in the DTOP (see Eq. (\ref{eqn:index})) at the critical times $t_{c,n}=(2n-1)t_c$. In Fig. \ref{fig:one}, we show a color-plot of the PGP as a function of $k$ and $t$ from which the critical times and the associated changes in the phase winding number $\nu_D$ representing our DTOP become visually clear (left panel). Moreover, Fig.~\ref{fig:one} displays the time dependence of both the DTOP $\nu_D(t)$ and the rate function $g(t)=-N^{-1} \log(|\mathcal{G}(t)|^2)=-\pi^{-1}\text{Re}\left(\log\int_0^\pi\text{d}k\,\left[ \lvert g_k\rvert^2 + \text{e}^{-2i\epsilon^f_k t}\lvert e_k\rvert^2\right]\right)$ which plays the role of a thermodynamic potential here and whose points of nonanalytic behavior define the DQPTs~\cite{Heyl2013a} (right panel). The DTOP indeed changes its value at every DQPT and uniquely characterizes the dynamical phase in between two DQPTs.

As a second benchmark, we consider a quench from $\vec d_i(k)=(0,0,1)$ to $\vec d_f(k)=(0,\sin(k),1+\cos(2k)+ \lambda \cos(k))$. This model is similar to the Kitaev chain studied before, but also includes a next-to-nearest neighbor hopping. For $0<\lambda<2$, this quench does not change the topological phase of the Hamiltonian since $\hat d_i(k_{\mathbb R})=\hat d_f(k_{\mathbb R})$.
Still, from Eq. (\ref{eqn:zerocond})
we find two critical momenta $k_c^{(1)}=\pi/2,~k_c^{(2)}= \arccos(- \lambda/2)$. At $k_c^{(1)}$, $\hat d_f(k)$ enters the northern hemisphere of the relative Bloch sphere and returns to the southern hemisphere at $k_c^{(2)}$, i.e., $\hat d_i(k)\cdot \hat d_f(k)<0$ for $k_c^{(1)}<k<k_c^{(2)}$. 
As a consequence, $\text{sgn}(s_{k_c^{(1)}})=-\text{sgn}(s_{k_c^{(2)}})=1$ and,  from Eq. (\ref{eqn:index}), we see that the change in the DTOP is opposite at the two critical momenta. In Fig. \ref{fig:two}, we show a color plot of the PGP for this quench (left panel) as well as the time dependence of the rate function $g(t)$ and the DTOP $\nu_D(t)$ (right panel). The DTOP changes at every DQPT, however, due to the competing $\Delta(t_{c,n}^{(i)})=(-1)^{i+1},~i=1,2$, its behavior is not monotonous as opposed to the first quench example.

\section{Relation to String order parameter}
\label{sec:stringorder}
It is natural to ask how the DTOP defined in Eq. \ref{eqn:dtop} is connected to the underlying equilibrium topology of the system. In this context, we show in the following numerical evidence that the dynamics of the DTOP can be linked to the decay of string order parameters. In thermal equilibrium, the string order parameter directly reflects the bulk topological properties. Focusing on the continuation of such bulk properties to non-equilibrium systems, our present study complements previous work reporting deviations from the  bulk-boundary correspondence out of thermal equilibrium~\cite{FosterPRB,FosterPRL,RigolPreprint,CaioPreprint}. There, surface states have been shown to develop dynamically even if the conventional bulk topological quantum numbers remain trivial~\cite{CaioPreprint}. 

\begin{figure}[tp]
\centering
\includegraphics[width=\columnwidth]{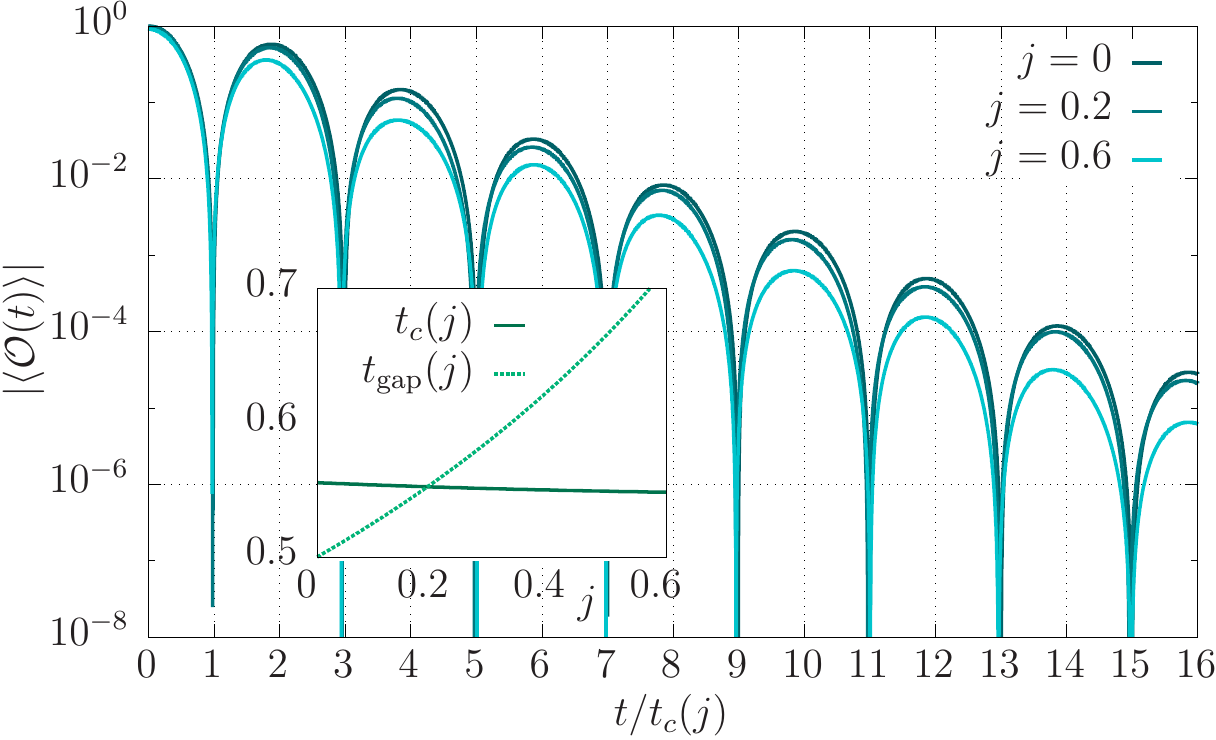}
\caption{\label{fig:three} Decay of the string order parameter for the same chemical potential quench as in Fig.~\ref{fig:one} but for a model also including a next-to-nearest neighbor hopping with amplitude $j$. The time axis is rescaled by $t^\ast$ where DQPTs occur and the DTOP changes its quantized value. The inset shows a comparison of the low-energy time scale $t_\mathrm{gap}(j) = \Delta(j)^{-1}$ with $\Delta(j)$ the gap, and $t_c(j)$. The simulations are done for a lattice system with up to $N=100$ sites, where we find that the data is converged with respect to system size.}
\end{figure}

Rather than exhibiting local order, topological phases are typically characterized by nonlocal properties. The topological phase in the Kitaev chain \cite{Kitaev2001} (see also first benchmark in Section \ref{sec:simulations}), is associated with a nonvanishing expectation value of the so called string order parameter
\be
      \mathcal{O}_{lm} =  \big( c_l+c_l^\dag \big) e^{i\pi \sum_{j=l}^{m-1} c_j^\dag c_j}  \big( c_m+c_m^\dag \big).
\ee
Here, $c_l$ with $l=1,\dots,N$ denotes the fermionic annihilation operator in its real-space representation. In Fig.~\ref{fig:three} we show numerically obtained dynamics $\mathcal{O}(t) = \lim_{|l-m|\to\infty}\langle \mathcal{O}_{lm}(t) \rangle$ for the string order parameter in the Kitaev chain for the same set of parameters as in Fig.~\ref{fig:one}, but for a slightly more general model including also a next-to-nearest neighbor hopping with strength $j$, i.e., $d^3(k) = \mu - \cos(k) - j\cos(2k)$. We have calculated the dynamics of $\mathcal{O}$ using the mapping onto Pfaffians~\cite{Barouch} which can be evaluated numerically very efficiently. For $j=0$ where the Kitaev chain can be mapped onto a transverse-field Ising model, the string order parameter maps to the spin-spin correlation function of the Ising order parameter whose dynamics has been studied previously~\cite{Calabrese2012,Heyl2013a}.

As one can see in Fig.~\ref{fig:three}, starting initially in the topological phase and quenching into the trivial phase, the string order parameter exhibits a damped oscillatory decay. The decay of the string order parameter is, of course, compatible with the topologically trivial structure of the final Hamiltonian. 
In the context of the DTOP, it is, however, of particular importance \emph{how} the string order decays. Specifically, the time scale for the oscillations on top of the exponential envelope coincides with $t_c(j)$ setting the critical times where DQPTs occur and the DTOP changes dynamically. To show this, we have rescaled the time axis by $t_c(j)$. After an initial transient regime where the time scale for the oscillations does not perfectly match $t_c(j)$ yet, in the asymptotic decay regime the agreement becomes very good. 
To further strengthen this suggested connection, we have also included an inset where we compare the emergent nonequilibrium time scale $t_c(j)$ with the conventional time scale $t_\mathrm{gap}(j) = \Delta^{-1}(j)$, $\Delta(j)$, i.e., the gap of the $j$-dependent the final Hamiltonian.  $t_c(j)$ we have extracted by solving Eq.~(\ref{eqn:zerocond}). While $t_\mathrm{gap}(j)$ increases with increasing $j$, we find that $t_c(j)$ decreases, thereby excluding the possibility of an accidental similarity of the two scales. This gives further evidence that the time scale associated with the string order parameter is not set by the gap but rather by the emergent nonequilibrium time scale $t_c(j)$ and thus is directly connected to the dynamics of the DTOP. 

\section{Concluding discussion}
\label{sec:discussion}
The DTOP $\nu_D$ defined here (see Eq. (\ref{eqn:dtop})) qualitatively distinguishes periods of time-evolution that are separated by DQPTs. As a truly dynamical quantity, the DTOP is fundamentally different from the conventional topological invariants that classify ground states of gapped Hamiltonians \cite{Schnyder2008,Kitaev2009,Ryu2010}. In particular, the conventional topological invariant for the time-dependent state $\lvert \psi(t)\rangle$ is a constant of motion in our present non-equilibrium setting \cite{Hickey2014,RigolPreprint,CaioPreprint}. However, there is an interesting interplay between equilibrium invariants  and the occurrence of DQPTs. In this context, it has been shown \cite{VanjaDoraPreprint} that a quench between topologically inequivalent Hamiltonians implies the presence of DQPTs, in agreement with the situation in our first benchmark example. However, DQPTs can also happen if the initial and the final Hamiltonian are topologically equivalent as in our second benchmark example. Remarkably, the structure of the DTOP is capable of resolving these different scenarios. From Eq.~(\ref{eqn:index}), we see that the DTOP behaves qualitatively different in these two cases (see comparison of right panels of Fig. \ref{fig:one} and Fig. \ref{fig:two}): If the Hamiltonian topology changes, there is an odd number of critical momenta giving rise to a change of the DTOP after one DQPT associated with each of the critical momenta. In contrast, for quenches between equivalent Hamiltonians, the sum over the changes of $\nu_D$ for all critical momenta is zero.

The construction of the DTOP relies on the two-banded character of the models we consider. For larger unit cells, pertaining in particular to the modelling of disordered systems, the change of the PGP between the real momenta is not necessarily quantized because the spectral PHS constraint in Eq.~(\ref{eqn:phs}) does not enforce Eq.~(\ref{eqn:hreal}). However, in many physical situations considering an effective model with two bands is a physically well justified approximation and the DTOP defined here is hence also expected to emerge in more complex systems. Furthermore, the general observation that Fisher zeros in real time lead to $\pi$ phase jumps in the PGP is valid beyond the specific construction of the DTOP. This may serve as a starting point for the generalization of our present construction to higher spatial dimensions in future work.

The dynamical order parameter identified in this work is of topological nature. A natural question is whether also \emph{local} dynamical order parameters can exist. Although a general answer is beyond the scope of this work, we can provide some intuitive insights. A dynamical analog of a diverging correlation length, giving rise to nonanalytic behavior of local observables in continuous equilibrium phase transitions, cannot develop dynamically in a finite period of time. This is due to fundamental causality constraints such as Lieb-Robinson bounds. Hence, we do not expect direct analogs of such phenomena at DQPTs.
  
We conclude by summarizing some recent experimental progress on the building blocks of a setup where DQPTs and the DTOP could be observed. Models similar to the Kitaev chain \cite{Kitaev2001}, see our benchmark examples, can be realized both in solid state systems~\cite{Mourik2012,Deng2012,Rokhinson2012,Das2012,Williams2012,Nadj-Perge2014,Lee2014} and potentially with cold atoms in optical lattices~\cite{Jiang2011,Nascimbene2013}. Instead of actually realizing a superfluid system described by a Bogoliubov de\,Gennes equation, experimental studies may resort to insulating band structures similar to the Su, Schrieffer, and Heeger model \cite{SSH} which also exhibits a formal PHS and for which the DTOP is analogously defined. While inducing nonequilibrium dynamics in solid state systems is challenging, with ultracold atoms quantum quenches have already been studied experimentally~\cite{Bloch2008,Polkovnikov2011}. Moreover, momentum-resolved phase differences of Bloch functions, needed for the DTOP, have recently been measured in terms of Berry phases~\cite{Atala2013,Aidelsburger2014,ChristofBerry}. In particular, the experimental techniques employed in Ref. \cite{ChristofBerry} can be directly employed to extract the PGP and to reconstruct the DTOP defined in the present work. Paving the way towards the general observation of DQPTs, a measurement scheme for Loschmidt echos $\mathcal{L}(t) = \lvert \mathcal{G}(t) \rvert^2$ has been introduced~\cite{Daley2012,Pichler2013}.

\section*{Acknowledgment} We acknowledge valuable discussions with Michael Tomka  and Sebastian Diehl. This work has been supported
by the Deutsche Akademie der Naturforscher Leopoldina
via Grant No. LPDS 2013-07 and LPDR 2015-01, as well as the ERC synergy grant UQUAM. M. H. thanks the Boston University Visitors Program for hospitality.

\bibliographystyle{apsrev}

\end{document}